\documentclass[12pt,a4paper]{article}
\usepackage{graphics}
\newcommand\as{\alpha_{\mathrm{S}}}
\newcommand\ycut{y_{\mathrm{cut}}}
\newcommand\yjoin{y_{\mathrm{join}}}
\newcommand\smfrac[2]{{\textstyle\frac{#1}{#2}}}

\def\ib#1#2#3{ibid.\ #1 (19#3) #2}
\def\np#1#2#3{Nucl.\ Phys.\ B#1 (19#3)~#2}
\def\pl#1#2#3{Phys.\ Lett.\ #1B (19#3) #2}

\def\zp#1#2#3{Zeit.\ Phys.\ C#1 (19#3) #2}

\def\be{\begin{equation}}
\def\ee{\end{equation}}
\def\ba{\begin{eqnarray}}
\def\ea{\end{eqnarray}}
\def\bann{\begin{eqnarray*}}
\def\eann{\end{eqnarray*}}
\def\benn{\begin{displaymath}}
\def\eenn{\end{displaymath}}
\def\nn{\nonumber}

\def\tq{\, \tilde{q}}

\def\n{\, {\cal N}}

\begin{document}

\begin{titlepage}
\begin{flushright}
  RAL-TR-98-041 \\ hep-ph/9805413
\end{flushright}
\par \vspace{10mm}
\begin{center}
{\Large \bf
The Jet Multiplicity as a Function of Thrust}
\end{center}
\par \vspace{2mm}
\begin{center}
{\bf D. J. Miller}\\
{and}\\
{\bf Michael H. Seymour}\\
\vspace{5mm}
{Rutherford Appleton Laboratory, Chilton,}\\
{Didcot, Oxfordshire.  OX11 0QX\@.  England.}
\end{center}
\par \vspace{2mm}
\begin{center} {\large \bf Abstract} \end{center}
\begin{quote} 
\pretolerance 1000
  We calculate the average multiplicity of jets in $e^+e^-$
  annihilation as a function of both the jet resolution parameter,
  $\ycut$, and the thrust $T$. Our result resums to all orders the
  leading and next-to-leading logarithms in $\ycut$ and $1-T$, and is
  exact up to second order in $\as$. This allows a comparison between
  the number of jets found using jet algorithms and the ability to
  distinguish different jet topologies via thrust.
\end{quote}
\vspace*{\fill}
\begin{flushleft}
  RAL-TR-98-041 \\ May 1998
\end{flushleft}
\end{titlepage}

\subsection*{1 Introduction}

The study of jets and their physics in $\mathrm{e^+e^-}$ annihilation
has added much to our understanding of perturbative QCD\@. Although
hadrons are the final state particles seen in such collisions, it is
quarks and gluons whose dynamics are described by perturbative QCD\@.
The definition of jets, via infrared and collinear safe jet clustering
algorithms, bridges the gap between the theoretically accessible
partons and the experimentally observed hadrons.

The multiplicity of jets in $\mathrm{e^+e^-}$ annihilation events
provides a good example of this. By contrasting the jet multiplicity
with the hadron multiplicity we may see some of the advantages of
using jets as our object of study. In particular, when predicting the
number of hadrons in the final state, one must invoke some
non-perturbative hadronization model at an unknown hadronization scale
$Q_0$. This results in an arbitrary overall normalisation, so that
only the energy variation of the hadron multiplicity can be calculated
within perturbative QCD\@. In contrast, by examining the jet
multiplicity, the hadronization scale is replaced by the known jet
resolution scale $Q_0=Q \sqrt{\ycut}$, and no arbitrary parameters are
introduced. Therefore, the jet multiplicity is fully calculable within
perturbative QCD, including its absolute value. In fact, since the jet
resolution scale is firmly within our control it can be varied at
will, effectively studying the energy dependence of the multiplicity
in a single experiment.

For large values of the jet resolution scale, the jet multiplicity is
reliably predicted by fixed-order perturbation theory. However, the
appearance of logarithms of $\ycut$ spoils the perturbative expansion
for finer jet resolutions and these logarithms must be summed to all
orders. With the advent of the $k_{\perp}$ (or Durham) jet finding
algorithm[\ref{cdotw}], it is possible to sum the leading and
next-to-leading logarithms of $\ycut$ to all orders,
making the jet multiplicity also reliable for small resolution scales[\ref{cdfw}]. This is the jet algorithm that we will use
throughout this paper, and it defines jets according to the following
iterative procedure:
\begin{itemize}
\item For each pair of particles calculate a separation defined by,
  \be y_{ij}=2 \frac{\mathrm{min}(E_i^2,E_j^2)}{Q^2}(1-\cos
  \theta_{ij}), \ee where $E_i$ and $E_j$ are the particles' energies
  and $\theta_{ij}$ is the angle between their momenta.
\item For the pair of particles with the smallest separation,
  $y_{ij}$, combine their momenta, $p_i$ and $p_j$, together to form a
  pseudo-particle of momentum $p_{ij}$. In the $E$-scheme, $p_{ij}=p_i +
  p_j$.
\item Repeat the above two steps until all the separations of
  particles or pseudo-particles are less than the jet resolution
  parameter $\ycut$. The remaining particles and pseudo-particles are
  then called jets.
\end{itemize}
This algorithm can be applied to form jets from either partons (for
theoretical considerations) or hadrons (for experimental analysis).
The difference between performing the algorithm at the partonic and
hadronic levels is expected to be suppressed by at least one power of
$1/Q$~[\ref{brwrev}].

The theoretical calculation and experimental analysis of event shapes
have also contributed greatly to our understanding of perturbative
QCD\@. In particular, the thrust and heavy jet mass of an event have
been extensively studied[\ref{cttw}] and show good agreement
with experiment[\ref{exptrev}]. It is interesting to calculate
the multiplicity of jets in $\mathrm{e^+e^-}$ annihilation, retaining
the jet kinematics so that the multiplicity can be expressed for
specific values of one of these event shapes. In this paper we
calculate the multiplicity of jets in $\mathrm{e^+e^-}$ annihilation
as a function of the thrust, summing all large logarithms to
next-to-leading logarithmic accuracy. This was done for the
multiplicity of hadrons in Ref.~[\ref{kt}]. The thrust for an event
is defined by,
\be
T= \mathrm{max} \left\{ \frac{\Sigma_i |\vec{P}_i
    \cdot \vec{n}|} {\Sigma_i |\vec{P}_i|} \right\},
\ee
where $\vec{n}$ is a unit vector chosen to maximise the
right-hand-side, and $\vec{P}_i$ are the three-momenta of the final
state partons or hadrons. Of course, our results can be easily applied
to other similar observables, such as the heavy jet mass.

The structure of the paper will be as follows. In section 2 we shall
consider the fixed-order contribution to the multiplicity up to ${\cal
  O}(\as^2)$. Then, in section 3, we will resum large logarithms of
$\tau \equiv 1-T$ and $\ycut$ to all orders in $\as$, and to
next-to-leading logarithmic accuracy, using the coherent branching
formalism. To avoid double counting, we will discuss the matching of
this result to fixed order in section 4, and finally, we present
numerical results in section 5.

\subsection*{2 Fixed Order}

We begin by defining the quantity $N(\tau,\ycut)$ as the multiplicity
of subjets resolved in the $k_\perp$ algorithm at a scale $\ycut$, in
events with thrust greater than $1-\tau$. It is generally given by,
\be N(\tau,\ycut)= \sum_{n=2}^{\infty} \frac{n \,
  \sigma_n(\tau,\ycut)}{\sigma_{\mathrm{tot}}}, \ee where
$\sigma_n(\tau,\ycut)$ is the cross section for events with thrust
greater than $1-\tau$ and exactly $n$ jets, and
$\sigma_{\mathrm{tot}}$ is the total cross section for the production
of hadrons. We will require this up to second order in the strong
coupling constant, $\as$. Defining $\sigma^{(m)}_n(\tau,\ycut)$ to be
the ${\cal{O}}(\as^m)$ contribution to $\sigma_n(\tau,\ycut)$, the
multiplicity to ${\cal O}(\as^2)$ is given by,
\[ \hspace{-13cm} N(\tau,\ycut) =  \]
\be \hspace{1cm}
   \frac{
  2 \, (\sigma_0 + \sigma^{(1)}_2(\tau,\ycut)+\sigma^{(2)}_2(\tau,\ycut)) +
  3 \, (\sigma^{(1)}_3(\tau,\ycut)+\sigma^{(2)}_3(\tau,\ycut)) +
  4 \, \sigma^{(2)}_4(\tau,\ycut)}
  {\sigma_0 + \frac{\as}{\pi} \, \sigma_0 +
  \left(\frac{\as}{\pi}\right)^2C\sigma_0},\phantom{(99)}
\ee
where $\sigma_0$ is the Born cross-section and $C$ is some number
that we will not eventually require. Truncating at ${\cal{O}}(\as^2)$,
we have:
\[  N(\tau,\ycut) =
  2\left[1-\frac{\as}{\pi}-C\left(\frac{\as}{\pi}\right)^2
    +\left(1-\frac{\as}{\pi}\right)\frac1{\sigma_0}\sigma^{(1)}_2(\tau,\ycut)
    +\frac1{\sigma_0}\sigma^{(2)}_2(\tau,\ycut)
    +\left(\frac{\as}{\pi}\right)^2
  \right] \]
\be
\hspace{1cm} +3\left(
    \left(1-\frac{\as}{\pi}\right)\frac1{\sigma_0}\sigma^{(1)}_3(\tau,\ycut)
    +\frac1{\sigma_0}\sigma^{(2)}_3(\tau,\ycut)\right)
  +4\frac1{\sigma_0}\sigma^{(2)}_4(\tau,\ycut).
\ee
Since at ${\cal O}(\as)$ every event has either 2 or 3 jets and at
${\cal O}(\as^2)$ every event has 2, 3 or 4 jets, we can write,
\ba
   \sigma^{(1)}(\tau) &=&
   \sigma^{(1)}_2(\tau,\ycut)+\sigma^{(1)}_3(\tau,\ycut), \\
   \sigma^{(2)}(\tau) &=&
   \sigma^{(2)}_2(\tau,\ycut)+\sigma^{(2)}_3(\tau,\ycut)
  +\sigma^{(2)}_4(\tau,\ycut).
\ea
This allows us to write the multiplicity in terms of more inclusive
quantities. Further simplification is made by rewriting our expression
in terms of the cross section for events with thrust less than
$1-\tau$, $\bar\sigma(\tau)$, using,
\ba
\sigma^{(1)}(\tau)+\bar\sigma^{(1)}(\tau)=\left(\frac{\as}{\pi}\right)
  \sigma_0, \\
\sigma^{(2)}(\tau)+\bar\sigma^{(2)}(\tau)=\left(\frac{\as}{\pi}\right)^2
  C \sigma_0.
\ea
This gives,
\ba
  N(\tau,\ycut) &=&
  2\left(1
    -\left(1-\frac{\as}{\pi}\right)\frac1{\sigma_0}\bar\sigma^{(1)}(\tau)
    -\frac1{\sigma_0}\bar\sigma^{(2)}(\tau)
  \right)
\nn \\ &&
  +\left(
    \left(1-\frac{\as}{\pi}\right)\frac1{\sigma_0}\sigma^{(1)}_3(\tau,\ycut)
    +\frac1{\sigma_0}\sigma^{(2)}_3(\tau,\ycut)\right)
  +2\frac1{\sigma_0}\sigma^{(2)}_4(\tau,\ycut)\phantom{(99)} \nn
\\&=&
  2\left(1-R(\tau)\right)+R_3(\tau,\ycut)+2R_4(\tau,\ycut),
\label{eq:secordmult}
\ea
where $R(\tau)$ is the ratio of the cross section for events with
thrust less than $1-\tau$ to the total cross section, and
$R_n(\tau,\ycut)$ is the ratio for events with $n$ jets and thrust greater 
than $1-\tau$,
evaluated to the appropriate order in $\as$, i.e.
\ba 
R(\tau) \equiv \frac{1}{\sigma_{\mathrm{tot}}} \bar\sigma(\tau), &&
R_n(\tau,\ycut) \equiv \frac{1}{\sigma_{\mathrm{tot}}} \sigma_n(\tau,\ycut). 
\ea

\subsubsection*{{\it 2.1 Zeroth Order}}

At zeroth order, ${\cal O}(\as^0)$, i.e.\ with no gluon emission, the
cross section is given by the production of a quark--antiquark pair,
emitted back-to-back in the centre of mass frame. The quark and
antiquark will always be placed in separate jets, so the jet
multiplicity will be two, and the thrust axis will fall along their
common axis, trivially giving a thrust of one. Therefore we have,
\begin{equation}
  N(\tau,\ycut)=2.
\end{equation}

\subsubsection*{{\it 2.2 First Order}}
We easily obtain the multiplicity to ${\cal O}(\as)$ by truncating
Eq.~(\ref{eq:secordmult}), giving,
\be N(\tau,\ycut) = 2\left(1-R(\tau)\right)+R_3(\tau,\ycut). \ee
Since neither of these cross sections involve the singular region of
thrust equal to 1, they can be calculated in four dimensions. At this
order, i.e.\ ${\cal O}(\as)$, the only contribution is from the
partonic process $\mathrm{e^+e^- \to q \bar q g}$.

For $R(\tau)$, the cross section for producing events with thrust
greater than $1-\tau$ divided by the total cross section to hadrons,
one obtains,
\ba
  R(\tau) &=&
 \frac{C_F\as}{2\pi}\Biggl\{2\log^2\tau+3\log\tau+\frac52-\frac{\pi^2}3
\nonumber\\&&\phantom{C_F\as}
  -6\tau\log\tau -4\log(1-\tau)\log\tau
  -6\tau -3(1-2\tau)\log(1-2\tau)
\nonumber\\&&\phantom{C_F\as}
  +4\mathrm{Li_2}\left(\frac{\tau}{1-\tau}\right)
  -\frac92\tau^2 +2\log^2(1-\tau)
  \Biggr\},
\ea
where only the first line contributes as $\tau\to0$.

Next we turn to $R_3(\tau,\ycut)$, the cross section for producing
three jets with thrust greater than $1-\tau$ divided by the total
cross section to hadrons. To this order,
\be R_3(\tau,\ycut) = \frac1{\sigma_0}\sigma^{(1)}_3(\tau,\ycut). \ee
The contributing partonic process is still $\mathrm{e^+e^- \to q \bar
  q g}$, but now we demand that the parton separations be greater than
$\ycut$ in order to maintain a three jet final state.

\vspace{0.2cm}

For $\ycut < \frac{2 \tau^2}{1-\tau}$ we have,
\ba
R_3(\tau,\ycut)&=&\frac{C_F \as}{2\pi} \left\{ \int_{1-\tau}^{1-\yjoin} dx_1
  \int_{1-x_1/2}^{x_1} dx_2 \right. \nn \\
&& \left. +\int_{1-\yjoin}^{1-\ycut} dx_1
 \int_{1-x_1/2}^{\frac{(1-x_1)(2-x_2)}{1+\ycut-x_1}} dx_2 \right\}
 \frac{2(x_1^3+x_2^3+(2-x_1-x_2)^3)}{(1-x_1)(1-x_2)(x_1+x_2-1)} \nn \\
&=& R_D(\ycut)-R(\tau), \ea
where,
\be \yjoin \equiv \frac{\sqrt{\ycut(8+\ycut)}-\ycut}{4}, \ee
and $R_D(\ycut)$ is the three--jet fraction in the $k_{\perp}$
algorithm at a scale $\ycut$, given in Ref.~[\ref{bs}].

\vspace{0.2cm}

For $\ycut > \frac{2 \tau^2}{1-\tau}$ this becomes,
\ba
R_3(\tau,\ycut)&=&\frac{C_F \as}{2\pi} \int_{1-\tau}^{1-\ycut} dx_1
 \int_{1-x_1/2}^{\frac{(1-x_1)(2-x_2)}{1+\ycut-x_1}} dx_2
 \frac{2(x_1^3+x_2^3+(2-x_1-x_2)^3)}{(1-x_1)(1-x_2)(x_1+x_2-1)} \nn \\
&=& \frac{C_F\as}{2\pi} \int_{1-\tau}^{1-\ycut} dx_1\frac{g(x_1,\ycut)}
{1-x_1}, \ea
with
\be
  g(x_1,\ycut) = \frac{2}{x_1}(2-3x_1(1-x_1))\log\frac{(1-x_1)(1-\ycut)}
  {\ycut-(1-x_1)^2}
  -3(2-x_1)^2\frac{1-\ycut-x_1}{1+\ycut-x_1}.
\ee
This integral has a closed analytic form, but does not give a very
compact expression, so we do not reproduce it here.

In order to match our resummed result to fixed order we must also
extract the logarithmic behaviour for $\tau,\ycut\ll1$. We obtain:
\begin{equation}
  \!R_3(\tau,\ycut) = \frac{C_F\as}{2\pi}
  \left\{ \begin{array}{ll}
      2\log^2\frac{\tau}{\ycut}-3\log\frac{\tau}{\ycut}+\log64
      &\ycut>2\tau^2,\\
      \log^2\frac1{\ycut}-2\log^2\frac1{\tau}
      -3\log\frac{\tau}{\ycut}+\log64+\log^22
      &\ycut<2\tau^2.
    \end{array}\right.
\end{equation}
Note that these two solutions are continuous at the matching point,
and that it can be shifted, with logarithmic accuracy, to:
\begin{equation}
  R_3(\tau,\ycut) = \frac{C_F\as}{2\pi}
  \left\{ \begin{array}{ll}
      2\log^2\frac{\tau}{\ycut}-3\log\frac{\tau}{\ycut}+\log64
      &\ycut>\tau^2,\\
      \log^2\frac1{\ycut}-2\log^2\frac1{\tau}
      -3\log\frac{\tau}{\ycut}+\log64
      &\ycut<\tau^2,
    \end{array}\right.
\end{equation}
or:
\begin{equation}
  R_3(\tau,\ycut) = \frac{C_F\as}{2\pi}
  \left\{ 2\log^2\frac{\tau}{\ycut}
    -3\log\frac{\tau}{\ycut}
    -\Theta(\tau^2-\ycut)\log^2\frac{\tau^2}{\ycut}
  \right\}.
\end{equation}

\subsubsection*{{\it 2.3 Second Order}}

Since we could make the first-order calculation in four dimensions, it
is possible to calculate the second-order terms using a standard Monte
Carlo NLO program like EVENT2[\ref{event2}].

Now Eq.~(\ref{eq:secordmult}) requires no modification and the
multiplicity is,
\be
  N(\tau,\ycut) = 2\left(1-R(\tau)\right)+R_3(\tau,\ycut)+2R_4(\tau,\ycut),
\ee
where $R(\tau)$ and $R_3(\tau,\ycut)$ should now be evaluated to NLO,
and $R_4(\tau,\ycut)$ to leading order.

Notice that at this order we can study the dependence on the
renormalization scale, since we have
\begin{equation}
  \label{eq:scaledep}
  R(\tau) = \frac{\as(\mu)}{2\pi}A(\tau) +
  \left(\frac{\as(\mu)}{2\pi}\right)^2
  \left(B(\tau)-\smfrac32C_FA(\tau)
    +\frac{b}{2}A(\tau)\log\frac{\mu^2}{Q^2}\right),
\end{equation}
with $b=\frac{11}3C_A-\frac23N_f$, and likewise for $R_3(\tau,\ycut)$.

\subsection*{3 All-Orders Summation of Large Logarithms}

In this section we present the all-orders resummation of all leading
($\as^n \log^m \tau \log^{2n-m} \ycut$) and next-to-leading ($\as^n
\log^m \tau \log^{2n-m-1} \ycut$ and $\as^n \log^{m-1} \tau
\log^{2n-m} \ycut$) logarithms, which appear for small $y_{cut}$
and/or $\tau$. These logarithms can be summed to all orders using the
coherent branching formalism.

To next-to-leading logarithmic accuracy, we have,
\be N(\tau,y_{cut})=2F_q(\tau Q^2,Q^2) N_q(\tau Q^2, Q^2;y_{cut} Q^2). \ee
Here, $F_q(k^2,Q^2)$ is the probability that a quark formed at scale Q
has a mass below~$k$, and $\log F_q$ has been calculated in
Ref.~[\ref{cttw}] to next-to-leading logarithmic accuracy. Here we
only require the expression for $F_q$ itself to next-to-leading
logarithmic accuracy, which is,
\be
\label{eq:Fq}
F_q(k^2,Q^2) = \exp \left\{ C_F \log \left( \frac{Q^2}{k^2} \right) f_1
 \left( \frac{\as(Q)}{4\pi} b \log \left( \frac{Q^2}{k^2} \right) \right)
 + \smfrac{3}{2} C_F f_2  \left( \frac{\as(Q)}{4\pi} b \log \left(
 \frac{Q^2}{k^2} \right) \right) \right\},
\ee
where,
\ba
f_1( \lambda ) &=& \frac{2}{b \lambda}
\left[(1-2 \lambda ) \log \left( \frac{1}{1-2 \lambda} \right)
-2(1- \lambda )\log \left( \frac{1}{1- \lambda} \right)
\right], \\
f_2( \lambda ) &=& \frac{2}{b} \log \left( \frac{1}{1- \lambda} \right).
\ea
Also, $N_q(k^2,Q^2;Q_0^2)$ is the contribution to the multiplicity in
a quark jet formed at a scale $Q$ and resolved at a scale $Q_0$, from
jet masses below $k$.

In order to derive $N_q(k^2,Q^2;Q_0^2)$, let us first consider the
quantity $n_q^a(k^2,Q^2;Q_0^2) \, dk^2$ which is the multiplicity of
partons of species $a$ in the jet, where the squared mass of the jet
lies between $k^2$ and $k^2+dk^2$. Clearly,
\be
N_q(k^2,Q^2;Q_0^2)= \sum_{a=\{q,g\}} \int_0^{k^2} dq^2 n_q^a(q^2,Q^2;Q_0^2).
\ee
This obeys an exclusive evolution equation given by, \vspace{0.5cm}
\[ \hspace{-6cm}
n_q^a(k^2,Q^2;Q_0^2) = \Delta(Q^2) \, \delta_{qa} \, \delta(k^2) \,
 \Theta(Q-Q_0) \]
\[ \hspace{2cm} + \Theta(Q-Q_0) \int_0^{Q^2} \frac{dq^2}{q^2}
 \frac{\Delta(Q^2)}{\Delta(q^2)} \int_0^1 dz \frac{\as(z(1-z)q)}{2\pi}
 P_{gq}(z) \int_0^{\infty} dk_q^2 \int_0^{\infty} dk_g^2 \]
\[ \hspace{-3cm} \Big( n_g^a(k_g^2,z^2q^2) f_q(k_q^2,(1-z)^2q^2) \,
 \Theta(zq-Q_0) \]
\[+ n_q^a(k_q^2,(1-z)^2q^2;Q_0^2) f_g(k_g^2,z^2q^2) \Theta((1-z)q-Q_0) \Big) \]
\be \hspace{-4cm}
\delta \left( k^2 - z(1-z)q^2 - \frac{k_g^2}{z}-\frac{k_q^2}{1-z} \right).
\ee

This has a simple physical interpretation. Firstly, it is possible
that the quark does not emit any radiation. Then only the original
quark will contribute to the multiplicity if it is sufficiently hard
(i.e.\ $Q>Q_0$). Clearly the resulting jet will be massless. This
gives rise to the first term. Alternatively, the quark may survive
until it reaches a scale $q$ before emitting a gluon of mass $k_g$ and
light-cone momentum fraction $z$ (and leaving the quark with mass
$k_q$ and light-cone momentum fraction $1-z$). Then the resolvable
partons originating from both quark and gluon must be counted. The
function $f_q(k^2,Q^2) \, dk^2$ in the above is the probability that a
quark formed at a scale $Q$ has a squared mass between $k^2$ and
$k^2+dk^2$, i.e.\ it is the derivative of $F_q(k^2,Q^2)$ with respect
to $k^2$.  $\Delta$ is the Sudakov form factor,
\be
\log \Delta(Q^2) = -\int_0^{Q^2} \frac{dq^2}{q^2} \int_0^1 dz P_{gq}(z)
\frac{\as(z(1-z)q)}{2\pi}.
\ee

These equations have only formal meaning, because the integrals are
divergent and the form factor is zero. To make them physically
meaningful, they can be calculated explicitly by imposing an infrared
cutoff $z(1-z)q> \epsilon$, or by working in $d=4-2\epsilon$
dimensions. They are however infrared finite, so the limit $\epsilon
\to 0$ can be taken smoothly. Thus we always imply such a procedure.

Differentiating with respect to $\log Q^2$ yields an inclusive
evolution equation:
\[ \hspace{-3cm} Q^2 \frac{d}{dQ^2} n_q^a(k^2,Q^2;Q_0^2) =
\int_0^1 dz \frac{\as(z(1-z)Q)}{2\pi} P_{gq}(z)
\left\{ \int_0^{\infty} dk_q^2 \int_0^{\infty} dk_g^2 \right. \]
 \vspace{-0.5cm}
\ba
&\Big(
n_g^a(k_g^2,z^2Q^2;Q_0^2) f_q(k_q^2,(1-z)^2Q^2) \Theta(z^2Q^2-Q_0^2) \nn \\
& \hspace{1.5cm}+ n_q^a(k_q^2,(1-z)^2Q^2;Q_0^2) f_g(k_g^2,z^2Q^2)
 \Theta((1-z)^2Q^2-Q_0^2)
\Big) \nn \\ & \hspace{2cm} \left.
\delta \left(k^2 - z(1-z)Q^2 - \frac{k_g^2}{z}-\frac{k_q^2}{1-z}\right)
- n_q^a(k^2,Q^2;Q_0^2) \Theta(Q-Q_0) \right\}.
\ea
The mass of the gluon will not significantly contribute to the mass of
the jet so to next-to-leading logarithmic accuracy it can be omitted
from the $\delta$-function. This $\delta$-function and the
normalisation conditions,
\ba
\int_0^{\infty} dk^2 f_q(k^2,Q^2)&=&1 ,  \\
\int_0^{\infty} dk^2 n_q^a(k^2,Q^2;Q_0^2)&=& \n_q^a(Q^2;Q_0^2),
\ea
can then be used to integrate out $k_q^2$ and $k_g^2$.  Here
$\n_q^a(Q^2;Q_0^2)$ is the multiplicity of partons of species $a$
inside a quark jet, where the quark is emitted at a scale $Q$ and the
partons are resolved at a scale $Q_0$ and no demand is made on the
mass of the jet. This has been derived in Ref.~[\ref{cdfw}],
\ba
\n_q^q(Q^2;Q_0^2)&=& \n^-(Q^2;Q_0^2)+\frac{8}{3} \frac{C_F}{C_A} \frac{N_f}{b}
 {\cal \widetilde{N}}(Q^2;Q_0^2), \\
\n_q^g(Q^2;Q_0^2)&=& \n^+(Q^2;Q_0^2)-\frac{8}{3} \frac{C_F}{C_A} \frac{N_f}{b}
 {\cal \widetilde{N}}(Q^2;Q_0^2),
\ea
where,
\ba
\n^+(Q^2;Q_0^2)&=& z_1 \left( \frac{z_0}{z_1} \right)^B
 \left[ I_{B+1}(z_1)K_{B}(z_0)+K_{B+1}(z_1)I_{B}(z_0) \right], \nn \\
\n^-(Q^2;Q_0^2)&=& \left( \frac{z_0}{z_1} \right)^{\frac{8}{3}
 \frac{C_F}{C_A} \frac{N_f}{b}} \nn \\
{\cal \widetilde{N}}(Q^2;Q_0^2)&=& \left( \frac{z_0}{z_1} \right)^B
 \left[ I_{B}(z_1)K_{B}(z_0)-K_{B}(z_1)I_{B}(z_0) \right].
\ea
and
\be
\label{eq:Bdef}
B=\frac{1}{b} \left( \frac{11}{3} C_A +\frac{2N_f}{3}-\frac{4C_FN_f}{3C_A}
 \right).
\ee
$I_{\nu}$ and $K_{\nu}$ are the modified Bessel functions and we have
used the notation of Ref.~[\ref{cdfw}] where,
\be
\label{eq:newvars}
z_1^2=\frac{32\pi C_A}{b^2 \as(Q)}, \quad \quad
z_0^2=\frac{32\pi C_A}{b^2 \as(Q_0)}.
\ee

Removal of these integrations gives,
\ba
\lefteqn{
Q^2 \frac{d}{dQ^2} n_q^a(k^2,Q^2;Q_0^2) =
\int_0^1 dz \frac{\as(z(1-z)Q)}{2\pi} P_{gq}(z) } & \nn \\
& \hspace{-0.5cm} \Big(
\n_g^a(z^2Q^2;Q_0^2) f_q((1-z)(k^2-z(1-z)Q^2),(1-z)^2q^2) \Theta(zQ-Q_0)
\Theta(k^2-z(1-z)Q^2) \nn \\
& + n_q^a((1-z)(k^2-z(1-z)Q^2),(1-z)^2Q^2;Q_0^2) \Theta((1-z)Q-Q_0)
\Theta(k^2-z(1-z)Q^2) \nn \\ & \hspace{-10cm} - n_q^a(k^2,Q^2;Q_0^2)
 \Theta(Q-Q_0) \Big).
\label{eq:incev}
\ea
The above evolution is dominated by the emission of soft gluons ($z
\approx 0$), so to next-to-leading logarithmic accuracy $z$ can be
replaced by zero in any smooth functions.  Since neither $f_q$ nor
$n_q^a$ are smooth functions (they contain $1/k^2$-like terms) such a
replacement cannot be made in their arguments. However, their
integrated distributions,
\ba
F_q(k^2,Q^2)&=&\int_0^{k^2} dq^2 f_q(q^2,Q^2), \\
N_q^a(k^2,Q^2;Q_0^2)&=&\int_0^{k^2} dq^2 n_q^a(q^2,Q^2;Q_0^2),
\ea
which are the functions we are really interested in, are smooth and
obey the same evolution equations as $f_q$ and $n_g^a$ but with
different boundary conditions.

Integrating Eq.~(\ref{eq:incev}) over $k^2$ and replacing $z$ with
zero where appropriate gives to the required accuracy,
\ba
Q^2 \frac{d}{dQ^2} N_q^a(k^2,Q^2;Q_0^2) &=&
\int_0^1 dz \frac{\as(zQ)}{2\pi} P_{gq}(z)  \nn \\
&& \Big(
\n_g^a(z^2Q^2;Q_0^2) F_q(k^2-zQ^2,Q^2) \Theta(zQ-Q_0)
\Theta(k^2-zQ^2) \nn \\
&& + N_q^a(k^2-zQ^2,Q^2;Q_0^2) \Theta(k^2-zQ^2)
\Theta(Q-Q_0) \nn \\ &&- N_q^a(k^2,Q^2;Q_0^2)\Theta(Q-Q_0) \Big).
\ea
Furthermore, $k^2-zQ^2$ can be replaced with $k^2$ in $F_q$ and
$N_q^a$ to next-to-leading logarithmic accuracy, giving the final
evolution equation in integral form,
\ba
\lefteqn{N_q^a(k^2,Q^2;Q_0^2) = \delta_{qa}} &&  \nn \\
&+& \int_0^{Q^2}\frac{dq^2}{q^2}
\int_0^1 dz \frac{\as(zq)}{2\pi} P_{gq}(z) F_q(k^2,q^2) \n_g^a(z^2q^2;Q_0^2)
\Theta(k^2-zq^2) \Theta(zq-Q_0) \nn \\
&-& \int_{k^2}^{Q^2} \frac{dq^2}{q^2}
\int_{k^2/q^2}^1 dz \frac{\as(zq)}{2\pi} P_{gq}(z) N_q^a(k^2,Q^2;Q_0^2).
\ea
This has a formal solution:
\[ \hspace{-2cm}
N_q^a(k^2,Q^2;Q_0^2) =\exp \left\{ - \int_{k^2}^{Q^2} \frac{dq^2}{q^2}
\int_{k^2/q^2}^1 dz \frac{\as(zq)}{2 \pi} P_{gq}(z) \right\}
\Big\{ \n_q^a(k^2;Q_0^2)
\]
\[ \hspace{1.2cm}+ \int_0^{Q^2} \frac{dq^2}{q^2}
\exp \left[\int_{k^2}^{q^2} \frac{d\tq^2}{\tq^2}
\int_{k^2/\tq^2}^1 d\bar{z} \frac{\as(\bar{z}\tq)}{\pi} P_{gq}(\bar{z})
 \right] F_q(k^2,q^2) \]
\be \hspace{3.5cm}
\times \int_0^1 dz \frac{\as(zq)}{2\pi} P_{gq}(z) \n_g^a(z^2q^2;Q_0^2)
 \Theta(k^2-zq^2) \Theta(zq-Q_0) \Big\}.
\ee

\noindent The integrals are considerably simplified by inserting the
unintegrated expression for~$F_q$~[\ref{cttw}],
\be
F_q(k^2,Q^2) = \exp \left\{ - \int_{k^2}^{Q^2} \frac{dq^2}{q^2}
\int_{k^2/q^2}^1 dz \frac{\as(zq)}{2 \pi} P_{gq}(z) \right\},
\ee
so that the expression for $N_q^a$ becomes,
\be
N_q^a(k^2,Q^2;Q_0^2) = F_q(k^2,Q^2)
\left\{\n_q^a(k^2;Q_0^2) + {\cal C}_q^a(k^2,Q^2;Q_0^2) \right\},
\ee
where,
\be
{\cal C}_q^a(k^2,Q^2;Q_0^2)=\int_{k^2}^{Q^2} \frac{dq^2}{q^2} \int_0^1 dz
\frac{\as(zq)}{2\pi} P_{gq}(z)
\n_g^a(z^2q^2;Q_0^2) \Theta(k^2-zq^2) \Theta(zq-Q_0).
\ee

The remaining integral can be done by making a change of variable from
$q^2$ to $\tq^2=z^2q^2$. To next-to-leading logarithmic accuracy,
\ba
{\cal C}_q^a(k^2,Q^2;Q_0^2) & = &
\int_0^{k^4/Q^2} \frac{d\tq^2}{\tq^2} \int_{\tq/Q}^{\tq/k} dz
\frac{\as(\tq)}{2\pi} P_{gq}(z) \n_g^a(\tq^2;Q_0^2) \Theta(\tq-Q_0) \nn \\
& + &
\int_{k^4/Q^2}^{k^2} \frac{d\tq^2}{\tq^2} \int_{\tq^2/Q^2}^{\tq/k} dz
\frac{\as(\tq)}{2\pi} P_{gq}(z) \n_g^a(\tq^2;Q_0^2) \Theta(\tq-Q_0) \nn \\
& = & 2C_F \int_0^{k^4/Q^2} \frac{d\tq^2}{\tq^2} \frac{\as(\tq)}{2\pi}
\n_g^a(\tq^2;Q_0^2) \log\left(\frac{Q}{k}\right) \Theta(\tq-Q_0) \nn \\
& + & 2C_F \int_{k^4/Q^2}^{k^2} \frac{d\tq^2}{\tq^2} \frac{\as(\tq)}{2\pi}
\n_g^a(\tq^2;Q_0^2) \log\left(\frac{k}{\tq}\right) \Theta(\tq-Q_0).
\ea
Making the $\Theta$-functions more explicit and rearranging gives,
\ba
{\cal C}_q^a(k^2,Q^2;Q_0^2) & = &
\Theta(k^2-QQ_0) 2C_F \int_{Q_0^2}^{k^4/Q^2} \frac{d\tq^2}{\tq^2}
 \frac{\as(\tq)}{2\pi} \n_g^a(\tq^2;Q_0^2) \log\left(\frac{Q}{k}\right) \nn \\
& + & \Theta(k^2-Q_0^2) 2C_F \int_{Q_0^2}^{k^2} \frac{d\tq^2}{\tq^2}
\frac{\as(\tq)}{2\pi} \n_g^a(\tq^2;Q_0^2) \log\left(\frac{k}{\tq}\right)\nn \\
& - & \Theta(k^2-QQ_0) 2C_F \int_{Q_0^2}^{k^4/Q^2} \frac{d\tq^2}{\tq^2}
\frac{\as(\tq)}{2\pi} \n_g^a(\tq^2;Q_0^2) \log\left(\frac{k}{\tq}\right)\nn \\
& = & C_F \left\{ I^a(k^2;Q_0^2)-I^a(k^4/Q^2;Q_0^2) \right\},
\ea
where
\be
I^a(k^2;Q_0^2)=\Theta(k^2-Q_0^2) \int_{Q_0^2}^{k^2} \frac{dq^2}{q^2}
 \frac{\as(q)}{2\pi} \n_g^a(q^2;Q_0^2) \log\left(\frac{k^2}{q^2}\right).
\ee

This integral is most easily done using the change of variables,
Eq.~(\ref{eq:newvars}), with the obvious addition,
\be
z_k^2=\frac{32\pi C_A}{b^2 \as(k)},
\ee
so that it becomes,
\be
I^a(k^2,Q_0^2)= \Theta (z_k-z_0) \frac{1}{2C_A} \int_{z_0}^{z_k} \frac{dz}{z}
\n^a_g(z;z_0)(z_k^2-z^2).
\ee
Using the results,
\ba
\int_{z_0}^{z_k} \frac{dz}{z} \n^+(z;z_0)(z_k^2-z^2)
&=& 2 \left( \n^+(z_k;z_0)-1 \right) + 4B {\cal \widetilde{N}}(z_k;z_0), \\
\int_{z_0}^{z_k} \frac{dz}{z} {\cal \widetilde{N}}(z;z_0)(z_k^2-z^2)&=&
2 {\cal \widetilde{N}}(z_k;z_0) - z_k^2/z_0^2+1,
\ea
we obtain,
\ba
I^g(k^2,Q_0^2)&=& \Theta (z_k-z_0) \frac{1}{C_A} \left[ \n^+(z_k;z_0)-1
+ 2B {\cal \widetilde{N}}(z_k;z_0) \right. \nn \\
&& \left. \quad - \frac{C}{2}
\left( 2 {\cal \widetilde{N}}(z_k;z_0) - z_k^2/z_0^2+1 \right) \right], \\
I^q(k^2,Q_0^2)&=& \Theta (z_k-z_0) \frac{C}{2 \, C_F}
\left( 2 {\cal \widetilde{N}}(z_k;z_0) - z_k^2/z_0^2+1 \right),
\ea
where the parameters $B$ and $C$ are given by Eq.~(\ref{eq:Bdef}) and
\be
\label{eq:Cdef}
C=\frac{8}{3} \frac{N_f}{b} \frac{C_F}{C_A}.
\ee

However, the quantity of interest is the multiplicity of all jets,
independently of their flavour, i.e.\ $N_q=N^q_q+N^g_q$. Therefore we
finally have,
\be
N_q(k^2,Q^2;Q_0^2)=F_q(k^2,Q^2) \left\{ \n_q(k^2;Q_0^2)+C_F \left( I(k^2,Q_0^2)
-I(k^4/Q^2,Q_0^2) \right) \right\}, \ee
with,
\ba
I(k^2,Q_0^2)=&\Theta (z_k-z_0) \frac{1}{C_A} \left[ \n^+(z_k;z_0)-1
+ 2B {\cal \widetilde{N}}(z_k;z_0) \nn \right.  \\ & \left. \quad + (B-1)
\left( 2 {\cal \widetilde{N}}(z_k;z_0) - z_k^2/z_0^2+1 \right) \right].
\ea
The above equation conforms with na{\"\i}ve expectations. One might
expect that the number of jets of mass below $k^2$ resulting from a
quark created at a scale $Q^2$ would be simply the probability of
finding a jet of mass below $k^2$ multiplied by the number of jets
within. Indeed, this is the first term of our expression for $N_q$,
and the na{\"\i}ve expectation requires only the addition of
next-to-leading logarithmic corrections.

For completeness, we also present the result for $N_g(k^2,Q^2,Q_0^2)$,
the multiplicity of partons found within a {\em gluon\/} of squared mass
below $k^2$. The derivation closely follows that of the quark case
above and will not be reproduced here. For the tagged multiplicities
we have,
\be
N_g^a(k^2,Q^2;Q_0^2)=F_g(k^2,Q^2) \left\{ \n_g^a(k^2;Q_0^2)
+C_A \left( I^a(k^2,Q_0^2) -I^a(k^4/Q^2,Q_0^2) \right) \right\},
\ee
Here $F_g(k^2,Q^2)$ is the integrated gluon jet mass distribution[\ref{cttw}], and takes the same form as $F_q$,
Eq.~(\ref{eq:Fq}), but with the coefficients of $f_1$ and $f_2$ now
being $C_A$ and $b/2$ respectively. Also $\n_g^a(k^2;Q_0^2)$ is the
multiplicity of parton species $a$ found within the gluon[\ref{cdfw}].

Notice that the function $I^a$ is identical to that in the quark case.
This is because the only contribution to these functions is from the
{\em singular\/} parts of the splitting kernels in the evolution
equation. The appropriate kernel for the quark is $P_{gq}(z) = 2C_F/z
+ {\cal O}(1)$ and for the gluon is $P_{gg}(z) = 2C_A/z + {\cal O}(1)$
(the other kernels $P_{qq}$ and $P_{qg}$ have no singular parts and
cannot contribute). The only difference between the two cases is the
colour factor, $C_A$ or $C_F$, which is reflected in the coefficient
of $I^a$.

Clearly, the untagged multiplicity is given by,
\be
N_g(k^2,Q^2;Q_0^2)=F_g(k^2,Q^2) \left\{ \n_g(k^2;Q_0^2)
+C_A \left( I(k^2,Q_0^2) -I(k^4/Q^2,Q_0^2) \right) \right\}.
\ee

\subsection*{4 Matching to Fixed Order}
Of course, the resummed result also contains part of the fixed-order
contribution, which has already been included. It must be matched to
fixed order so that we do not over-count. This is done by expanding
the resummed result to next-to-leading order in $\as$, and
subtracting the offending piece from our result. Expanding each term
individually, we obtain:
\ba
  F_q(e^{-L}Q^2,Q^2) &\approx& 1-
  \frac{\as}{2\pi}\left(C_FL^2-\smfrac32C_FL\right) \nn \\ &&+
  \left(\frac{\as}{2\pi}\right)^2
  \left(\smfrac12C_F^2L^4-[\smfrac12bC_F+\smfrac32C_F^2]L^3
    +[\smfrac38bC_F+\smfrac98C_F^2]L^2\right),
\ea
where $\as=\as(Q)$,
\begin{eqnarray}
  {\cal{N}}_q(k^2;e^{-l}k^2) &\approx& 1+
  \frac{\as}{2\pi}\left(\smfrac12C_Fl^2-\smfrac32C_Fl\right)+
  \left(\frac{\as}{2\pi}\right)^2
  \Biggl(\smfrac1{24}C_FC_Al^4-\smfrac1{12}C_F[3C_A-b]l^3
\nonumber\\&&
  +\frac{2(C_A-C_F)C_FT_RN_f(8C_FT_RN_f-2C_AT_RN_f-C_A^2)}{9C_A^3}
  l^2\Biggr),
\end{eqnarray}
where $\as=\as(k)$, and finally,
\ba
  \lefteqn{I(k^2;e^{-l}k^2) \approx
  \frac{\as}{2\pi}\left(\smfrac12l^2\right)} &&  \nn \\ &&+
  \left(\frac{\as}{2\pi}\right)^2
  \Biggl(\smfrac1{24}C_Al^4+\smfrac1{12}bl^3
  -\frac{2(C_A-C_F)T_RN_f(11C_A^2-4C_FT_RN_f)}{9C_A^3}
  l^2\Biggr),
\ea
where again $\as=\as(k)$.

Note that, in the above, the coefficients of $\as^2L^2$ and $\as^2l^2$
are not believed to be correct, since they are only
next-to-next-to-leading logarithms. Nevertheless, they must be
considered when performing the matching since the fixed-order result
is exact to order $\as^2$.  It only remains to write the various $\as$
values in terms of $\as(Q)$:
\begin{eqnarray}
  \frac{\as(k)}{2\pi} &\approx& \frac{\as}{2\pi} +
  \smfrac12b\left(\frac{\as}{2\pi}\right)^2\log\frac{Q^2}{k^2},\\
  \frac{\as(k^2/Q)}{2\pi} &\approx& \frac{\as}{2\pi} +
  b\left(\frac{\as}{2\pi}\right)^2\log\frac{Q^2}{k^2}.
\end{eqnarray}

The result of this expansion is most easily written in the form of
Eqs.~(\ref{eq:secordmult}) and~(\ref{eq:scaledep}), with the
additional notation:
\ba
\widetilde{R}(\tau,\ycut) &=& R_3(\tau,\ycut)+2R_4(\tau,\ycut) \nn \\
&=& \frac{\as(Q)}{2\pi}\widetilde{A}(\tau,\ycut) +
  \left(\frac{\as(Q)}{2\pi}\right)^2
  \left(\widetilde{B}(\tau,\ycut)-\smfrac32C_F\widetilde{A}(\tau,\ycut)\right).
\ea
Then,
\ba
  A(e^{-L}) &=& 2C_FL^2-3C_FL, \\
  B(e^{-L}) &=& -2C_F^2L^4+C_F(6C_F+b)L^3
  -C_F(\smfrac34b+\smfrac32C_F)L^2-\smfrac92C_F^2L,\\
  \widetilde{A}(e^{-L},e^{-l-L}) &=&
  (2C_Fl^2-3C_Fl)\Theta(l)-C_F(l-L)^2\Theta(l-L),\\
  \widetilde{B}(e^{-L},e^{-l-L}) &=&
  \Biggl[(\smfrac12bL-(2C_FL^2-3C_FL)+\smfrac32C_F)(2C_Fl^2-3C_Fl) \nn \\
  &&+\smfrac16C_FC_Al^4+(\smfrac13b-\smfrac12C_A)C_Fl^3 \nn \\
  &&-\frac{8(C_A-C_F)C_FT_RN_f(6C_A^2+C_AT_RN_f-6C_FT_RN_f)l^2}{9C_A^3}
  \Biggr]\Theta(l) \nn \\
  &&-\Biggl[(bL-(2C_FL^2-3C_FL)+\smfrac32C_F)C_F(l-L)^2 \nn \\
  &&+\smfrac1{12}C_FC_A(l-L)^4+\smfrac16bC_F(l-L)^3 \nn \\
  &&-\frac{4(C_A-C_F)C_FT_RN_f(11C_A^2-4C_FT_RN_f)(l-L)^2}{9C_A^3}
  \Biggr]\Theta(l-L).\nn \\
\ea
These have been checked by solving the evolution equation iteratively,
verifying that the claimed solution does actually satisfy the
evolution equation, at least to second order in~$\as$.

\subsection*{5 Numerical Results}
\begin{figure}\vspace*{-1ex}
  \centerline{%
    \resizebox{!}{8cm}{\includegraphics{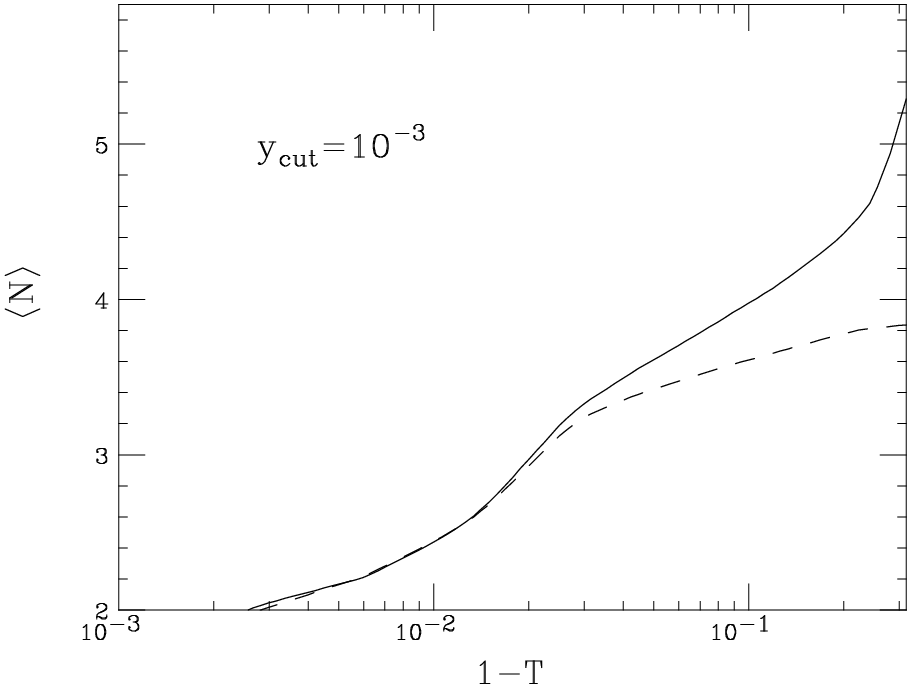}}
    }\vspace*{-2ex}
  \caption[]{The mean number of jets in $e^+e^-$ annihilation as a
    function of thrust, using the ${\cal{O}}(\as^2)$ result alone
    (dashed) and matched with the resummed result (solid).}
  \label{N(T)}
\end{figure}

Combining the fixed-order and resummed results, we obtain the
contribution to the mean jet multiplicity from events with thrust
between $T$ and $T+dT$, $n(1-T,\ycut)dT$.  To obtain the mean jet
multiplicity as a function of thrust we normalise this to the number
of events in the same range, $r(1-T)dT$, to give
\begin{equation}
  \langle N \rangle(T) = \frac{n(1-T,\ycut)}{r(1-T)},
\end{equation}
shown in Fig.~\ref{N(T)} for a representative value of
$\ycut=10^{-3}$.
We fix $\mu=\sqrt{s}=m_{\scriptscriptstyle{Z}}$ and
$\as(m_{\scriptscriptstyle{Z}})=0.120$.  We see that for $1-T\gg\ycut$,
the resummation is extremely important.  For $1-T\sim\ycut$, the
physical threshold, the NLO corrections are negative and even after
resumming large logarithms to all orders, give an unphysical result,
with the mean number of jets falling below 2.  This region is anyway
outside perturbative control, since it corresponds to
$k^2=s(1-T)\sim\Lambda\sqrt{s}$.

\begin{figure}\vspace*{-1ex}
  \centerline{%
    \resizebox{!}{8cm}{\includegraphics{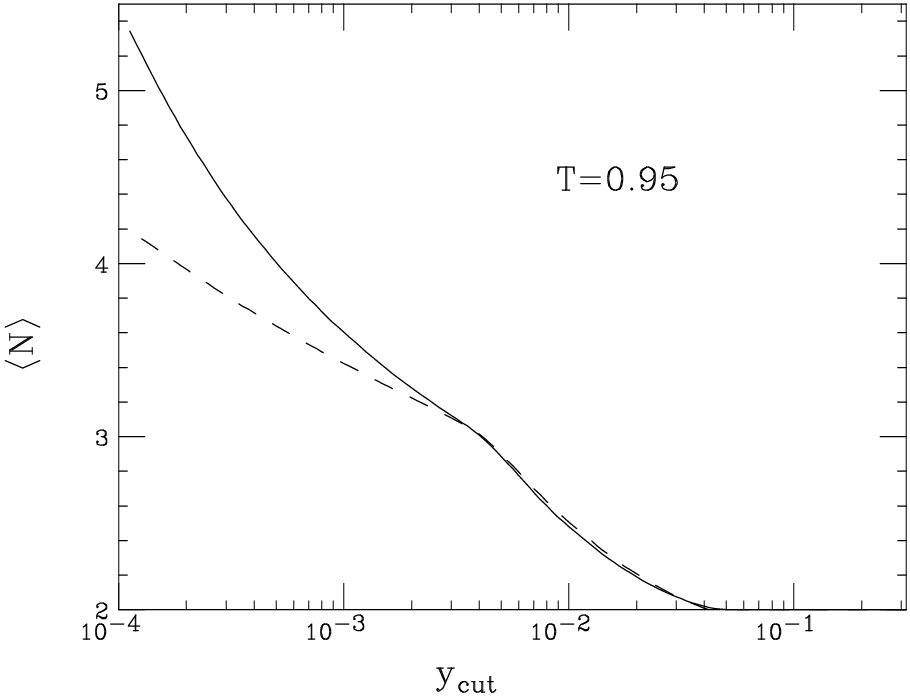}}
    }\vspace*{-2ex}
  \caption[]{The mean number of jets in $e^+e^-$ annihilation as a
    function of $\ycut$ for fixed thrust, using the ${\cal{O}}(\as^2)$
    result alone (dashed) and matched with the resummed result (solid).}
  \label{N(y)}
\end{figure}

In Fig.~\ref{N(y)} we show the result for fixed thrust $T=0.95$ as a
function of $\ycut$.
The resummation is again seen to be important for $\ycut\ll1-T$.

\begin{figure}\vspace*{-1ex}
  \centerline{%
    \resizebox{!}{8cm}{\includegraphics{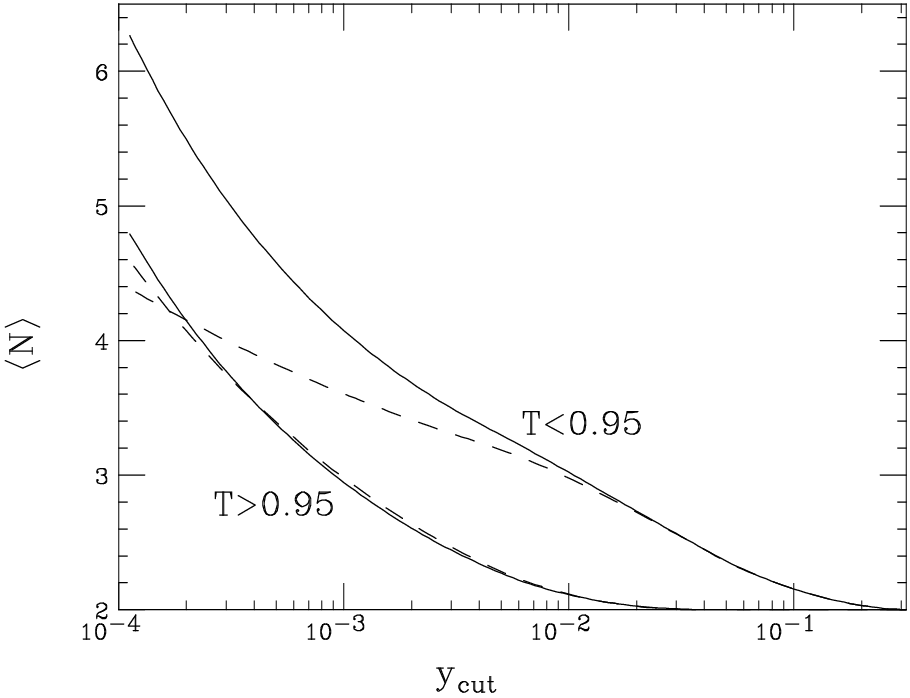}}
    }\vspace*{-2ex}
  \caption[]{The mean number of jets in $e^+e^-$ annihilation as a
    function of $\ycut$ for events with thrust above and below a fixed
    cut, using the ${\cal{O}}(\as^2)$ result alone (dashed) and matched
    with the resummed result (solid).}
  \label{two-three}
\end{figure}

The thrust can be used to separate two-jet from three-jet events.  In
Fig.~\ref{two-three} we show the multiplicity of jets in each sample.
These are defined by
\begin{equation}
  \label{T>t}
  \langle N \rangle(T>t) = \frac{N(1-t,\ycut)}{R(1-t)},
\end{equation}
and
\begin{equation}
  \label{T<t}
  \langle N \rangle(T<t) = \frac{N(\ycut)-N(1-t,\ycut)}{1-R(1-t)},
\end{equation}
respectively.  They can be directly compared with the results of
Ref.~[\ref{cdfw2}], where the separation into two- and three-jet
events was made using the same jet algorithm as the one in which the
jets were counted~--- the $k_\perp$ algorithm.  Note that for small
$\ycut$ the ${\cal{O}}(\as^2)$ results appear to be unphysical with
the multiplicity in two-jet events being higher than for three-jet
events.  This is somewhat misleading as they are effectively
calculated to different orders owing to the different denominators in
Eqs.~(\ref{T>t}) and~(\ref{T<t}),
\begin{eqnarray}
  \langle N \rangle(T>t) &\sim&
  \frac{2+{\cal{O}}(\as)+{\cal{O}}(\as^2)}{1+{\cal{O}}(\as)+{\cal{O}}(\as^2)}
  \sim 2+{\cal{O}}(\as)+{\cal{O}}(\as^2),
\\
  \langle N \rangle(T<t) &\sim&
  \frac{{\cal{O}}(\as)+{\cal{O}}(\as^2)}{{\cal{O}}(\as)+{\cal{O}}(\as^2)}
  \sim 3+{\cal{O}}(\as).
\end{eqnarray}
The all-orders results are physically behaved, because the first
uncalculated term in either result is suppressed by at least two orders
of $\as$ and two powers of $\log(\ycut)$.

One might na{\"\i}vely expect that the number of jets in a three-jet
event be larger than in a two-jet event by a factor $(2C_F+C_A)/(2C_F)
\approx 2$, which is certainly not the case in Fig.~\ref{two-three}
where for small $\ycut$ it is around 1.3.  This is comparable to the
ratio found in the $k_\perp$ algorithm of around 1.4 and, as discussed
in Ref.~[\ref{cdfw2}] is because the form factor suppression for the
gluon jet is always stronger than for the quark jets.

\begin{figure}\vspace*{-1ex}
  \centerline{%
    \resizebox{!}{8cm}{\includegraphics{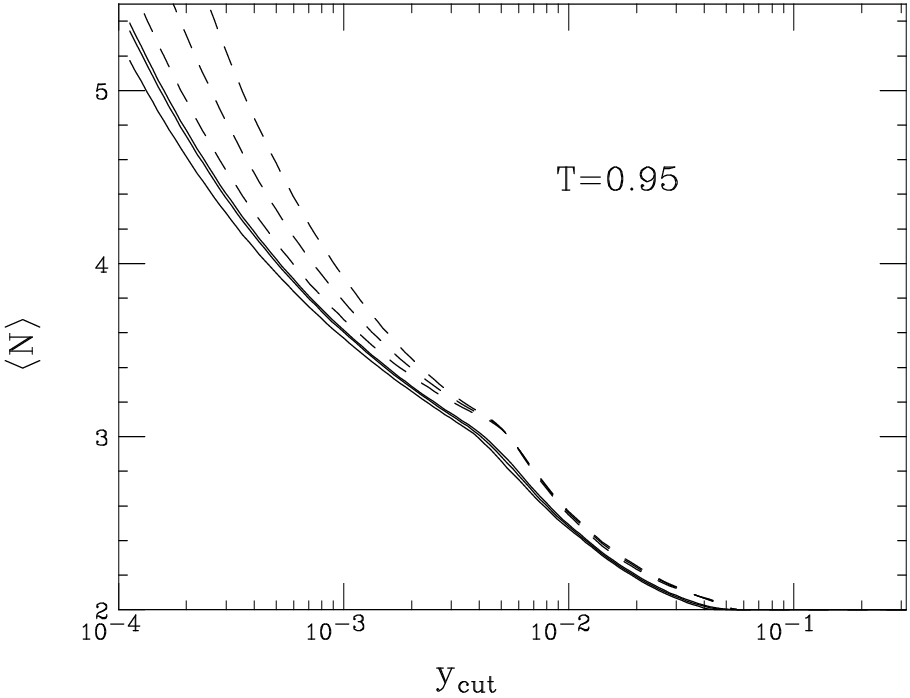}}
    }\vspace*{-2ex}
  \caption[]{The mean number of jets in $e^+e^-$ annihilation as a
    function of $\ycut$ for fixed thrust, using the resummed result
    matched with the ${\cal{O}}(\as)$ (dashed) and ${\cal{O}}(\as^2)$
    (solid) results, for renormalization scale choices $\mu=\sqrt{s}/2$,
    $\sqrt{s}$ and $2\sqrt{s}$.}
  \label{mu}
\end{figure}

Finally, in Fig.~\ref{mu} we show the variation of the result at fixed
thrust under changes of renormalization scale between $\sqrt{s}/2$ and
$2\sqrt{s}$.
We see that the ${\cal{O}}(\as^2)$ result matched with the all-orders
resummation is considerably more stable than the ${\cal{O}}(\as)$ plus
resummed result.

\newpage

\subsection*{References}

\begin{enumerate}
\item\label{cdotw}
S. Catani, Yu.L. Dokshitzer, M. Olssen, G. Turnock and B.R. Webber,
\np{269}{432}{91}
\item\label{cdfw}
S. Catani, Yu.L. Dokshitzer, F. Fiorani and B.R. Webber, \np{377}{445}{92}
\item\label{brwrev}
B.R. Webber `Renormalon Phenomena in Jets and Hard Processes', talk
given at 27th International Symposium on Multiparticle Dynamics
(ISMD 97), Frascati, Italy, 8--12 Sep 1997, hep-ph/9712236
\item\label{cttw}
S. Catani, L. Trentadue, G. Turnock and B.R. Webber, \np{407}{3}{93}
\item\label{exptrev}
See for example M. Schmelling, Phys.\ Scr.\ 51 (1995) 683 and references
therein
\item\label{kt}
K. Kimura and K. Tesima, \zp{62}{471}{94}
\item\label{bs}
N. Brown and W. J. Stirling, \zp{53}{629}{92}
\item\label{event2}
S. Catani and M.H. Seymour, \pl{378}{287}{96};
\np{485}{291}{97} and erratum \ib{510}{503}{97}
\item\label{cdfw2}
S. Catani, Yu.L. Dokshitzer, F. Fiorani and B.R. Webber, \np{383}{419}{92}
\end{enumerate}

\end{document}